\documentclass[12pt,a4paper]{article}
\usepackage{graphics}
\begin{document}
\,\,\,\,

\centerline{\bf GINZBURG-LANDAU THEORY OF VORTEX PHASE DIAGRAM}
\centerline{\bf IN LAYERED TYPE II SUPERCONDUCTOR}
\vspace{15mm}

\centerline{\bf RYUSUKE IKEDA}
\vspace{8mm}

\centerline{Department of Physics, Kyoto University, Kyoto 606-8502,
JAPAN}
\vspace{20mm}

\leftline{\bf 1. INTRODUCTION}
\vspace{5mm}
Various macroscopic phenomena seen in high temperature superconductors
(HTSC)
under nonzero magnetic fields have inspired much interest in studying
the phase diagram of vortex states of three dimensional (3D) layered
type II superconductors. Experimentally, thermodynamic and transport
phenomena have been
intensively examined mainly in YBCO [1-3] and BSCCO [4]. The resistive
broadening [5] in the so-called vortex liquid regime below the crossover
line $H_{c2}(T)$ is common to these materials with strong fluctuation
effect and is well described in terms of the non-Gaussian
superconducting fluctuation theory[6].
Since, on approaching a phase transition, macroscopic behaviors
tend to become insensitive to microscopic details of each material, one
expects that the phase diagram itself should be qualitatively the same
between these materials. However, recent
data [7-10] in YBCO seem to have shown some details of transition lines
in real systems with pinning disorders which have been unseen in BSCCO.
It
is important to find how such phenomena {\it apparently} dependent on
the
materials are explained within a single GL theory.

Theoretically, the phase diagram of vortex states has been tackled from
two
different points of view. In the literature [11-13] based on the elastic
theory for the vortex systems with pinning disorder, one first starts
from low fields
and low temperatures and hence, works in the London (phase-only) limit
of the
GL model. Consequently, a glassy solid phase, named Bragg-glass (BrG)
phase, was proposed as the ground state in cleaner real systems. Then, a
melting line
of BrG phase is estimated in terms of some kind of Lindemann criterion
and is usually identified with a simultaneous destruction of positional
and glass
(superconducting) orders. Since the positional ordering of field-induced
vortices in clean bulk systems is believed to occur through a first
order transition [14], however, this picture of glass ordering [11-13]
does not permit a continuous disappearance of linear resistance and
hence, is incompatible with transport phenomena in fields lower than a
lower critical point of clean YBCO samples [7-9] and in systems with a
continuous glass transition [15] due to strong correlated (line-like)
disorder.

By contrast, the approach [16-19]
from higher temperatures describes the vanishing of
linear resistance according to a superconducting glass ordering proposed
by Fisher et al. [20] for homogeneous (nongranular) type II
superconductors in nonzero fields. In layered materials, this ordering
occurs when the glass susceptibility, which is the spatial average of
correlation function [20]
$$
G_G(md, {\bf R})=d \sum_j \int d^2r {\overline {|< \, \psi^*_j({\bf
r}) \psi_{j+m}({\bf r}+{\bf R}) \, >|^2}} \eqno(1.1)
$$
expressed in terms of the pair-field (superconducting order parameter)
$\psi_j({\bf r})$ at $j$-th layer, becomes divergent. In eq.(1.1), the
angular bracket and the overbar denote, respectively, the thermal and
the random averages, and $d$ is the interlayer spacing. This approach
can be formulated [17-19] as a natural extension of the nonGaussian
superconducting fluctuation theory [6] to lower temperature at which the
vortex pinning due to structural disorder is not negligible {\it even in
clean systems}. Previously, this approach was criticized in a review
paper [21] because it was not easy [16] for this approach to justify the
phenomenological guess [20,21]
that, in thermodynamic limit, the vortex solid in the pinning-free case
with nonzero vortex flow resistance should
be replaced in clean limit of real systems by a vortex glass with zero
linear
resistance. This obstacle was overcome [17,22] at least in high field
case by combining properties of the pinning-free Abrikosov solid in 2D
limit with the framework of vortex glass fluctuation based on eq.(1.1).
Further, since eq.(1.1) is an expression independent of the detail of
random average, this approach is easily
extended [19]
for describing continuous glass transitions [15] induced by correlated
disorder. On the other hand, it is unclear at present to what extent
this approach can be extended into the resulting glass phases.

In this article, a theoretical development on the vortex glass
transitions based on eq.(1.1) is reviewed. Consistently with eq.(1.1),
the GL model for the layered system, i.e., the Lawrence-Doniach (LD)
model [23], will be used throughout this paper:
$$
{\cal H}_{\rm LD}=d \sum_j \int d^2r \biggl[ \, \biggl({T \over
{T_{c0}}}-1\biggr) |\psi_j|^2 + \xi_0^2 \biggl|\biggl(-{\rm
i}\nabla_\perp + {{2\pi} \over {\phi_0}} {\bf A}_{\perp, j} \biggr)
\psi_j \biggr|^2 $$
$$
+ \Gamma^{-1} \biggl({{\xi_0} \over d}\biggr)^2 \biggl|\psi_j -
\psi_{j+1}
\exp{\biggl({\rm i} {{2\pi d {\delta A_\parallel}} \over {\phi_0}}
\biggr)}
\biggr|^2 + {b \over 2} |\psi_j|^4 \biggr] \eqno(1.2)
$$
with disorder terms
$$
{\cal H}_{rp}=d \sum_l \int d^2r \biggl[ \, u_l({\bf r})
|\psi_l({\bf r})|^2 + f_l({\bf r}) \xi_0^2 (\nabla \times {\bf
j}_l)_\parallel \biggr], \eqno(1.3)
$$
where $\xi_0$ the in-plane coherence length, $\phi_0$ the flux quantum,
$b > 0$, $\Gamma$ the mass anisotropy, and the vector indices $\perp$
and $\parallel$ imply the directions, respectively, parallel and
perpendicular to the layer
plane. Throughout this paper, internal gauge fluctuations except the
external disturbance $\delta {\bf A}$
are neglected by focusing on the type II limit so that the applied field
is given by ${\rm curl} {\bf A}_{\rm ext}$ where ${\bf A}_{\rm ext}
= {\bf A} - \delta {\bf A}$. In eq.(1.3), ${\bf j}_l=\psi^*_l
( - {\rm i}\nabla + 2 \pi {\bf A}_{\rm ext}/\phi_0) \psi_l + {\rm
c.c.}$, and
the structural disorder is described by a random potential
expressing $T_{c0}$-variations, $u_j({\bf r})$, and a randomness of
flux $f_j({\bf r})$ [18,19]. We note that, in the phase-only limit
($|\psi|$ const.), the second term of eq.(1.3) expresses the pinning
of
vortex cores [11,12], while its first term becomes negligible.

This paper is organized as follows. In $\S 2$, the phase diagram of real
systems in the case with ${\bf B} \perp$ layers and, primarily with only
point (uncorrelated) disorder, is discussed according to recent works
[17-19, 24], which are
based on the theoretical findings that, in the pinning-free case, the
vortex
liquid region of the normal
metal phase discontinuously freezes to change into the Abrikosov
vortex solid with long-ranged positional order and quasi long-ranged
[25] (conventional) phase coherence and that, in real systems with
pinning disorder, a static glass ordering defined using eq.(1.1) occurs
while the conventional phase coherence remains short-ranged because of a
disorder-induced partial destruction [20,21] of positional long-ranged
order. The above second statement may be subtle if the BrG phase [11-13]
will occur just below the first order freezing of the vortex liquid. In
$\S 3$, the phase diagram in the case ${\bf B} \parallel $ layers of
real systems with point disorder is discussed by applying [26,27] the
treatment sketched in $\S 2$ to this case and is briefly compared with
existing data [28,29]. In $\S 4$, a relevance of results in $\S 2$ to
BSCCO in low fields perpendicular to the layers is discussed together
with an issue of
Hall conductivity near glass transitions.

\vspace{8mm}
\leftline{\bf 2. PHASE DIAGRAM IN FIELDS PERPENDICULAR TO LAYERS}
\vspace{5mm}

First, the model (1.2) will be rewritten within the subspace of the
lowest Landau level (LLL) of the $\psi$-fluctuation by invoking a high
field approximation
valid far from a critical region of the normal-Meissner transition at
$T_{c0}$. Further, the 2D case will be considered [22] for a while for
our
convenience of description. Using a Laudau-gauge and
expressing $\psi$ as $\psi({\bf r})=\sum_p \phi_p
u_p({\bf r})$ in terms of the LLL eigenfunction $u_p$, the model ${\cal
H} \equiv {\cal H}_{\rm LD} + {\cal H}_{rp}$ takes the form
$$
{\cal H}=\sum_p \, \mu_0 \, |\phi_p|^2 + \sum_{\bf k} \biggl({b
\over {2 \xi_0^2 d}} \, v_{\bf k} \, |{\tilde \rho}_{\bf k}|^2 +
(u_{-{\bf k}} + f_{-{\bf k}} \, {\bf k}^2 \xi_0^2 ) \, v^{1/2}_{\bf k}
{\tilde \rho}_{\bf k} \biggr), \eqno(2.1)
$$
where $h=2 \pi \xi_0^2 B/\phi_0$, $\mu_0=-1+h+T/T_{c0}$,
$B$ the magnitude of the applied field,
$v_{\bf k}=\exp(-{\bf k}^2/(2 h))$, and
$$
{\tilde \rho}_{\bf k}={{\xi_0} \over L} \sum_p \exp({\rm i} p k_1/h)
\,
\phi^*_{p-k_2/2} \, \phi_{p + k_2/2} \eqno(2.2)
$$
with linear system size $L$. Since the characteristic microscopic length
of a vortex state is the magnetic length $r_B \equiv \sqrt{\phi_0/(2 \pi
B)}$, the factor $k^2$ of the random-flux term implies that the random
potential $f$ is accompanied by the factor $h$ and hence that the
pinning disorder will be enhanced
with increasing field. This trend is also valid in the phase-only model
with no $u$-potential term and will be valid in general at least in type
II limit [20,21]. The Gaussian ensemble for the $u$-potential will be
assumed: ${\overline {u({\bf r}) \, u({\bf r}')}}=\Delta
\delta^{(2)} ({\bf r} - {\bf r}')$.
For just simplicity of our presentation, the $f$-potential term will be
omitted for a while.

First, let us consider, as typical quantities appearing even in the
pinning-free case, the pairing entropy density (pair-field propagator)
$$
{\overline {< |\phi_p|^2 >}}=N_v^{-1} \int d^2r {\overline {<
|\psi|^2 >}} \eqno(2.3)
$$
and the Abrikosov factor
$$
\beta_A={{2 \pi} \over h} N_v \biggl( \int d^2r {\overline {<
|\psi|^2 >}} \biggr)^{-2} {\int d^2r {\overline {< |\psi|^4 >}}},
\eqno(2.4)$$
where $N_v=L^2/(2 \pi r_B^2)$ is the number of vortices. The former
satisfies a Dyson equation
$$
\mu={{k_{\rm B} T} \over {\overline {< |\phi_p|^2 >}}}=\mu_0 +
\mu \, x \,
(\beta_A - \Delta_{\rm eff}(T) (\beta_A - 1)), \eqno(2.5)$$
where $x=k_{\rm B} T \, b \, h/(2 \pi \xi_0^2 d \mu^2)$, and
$\Delta_{\rm eff}(T)=\Delta d \xi_0^2 /k_{\rm B} T b$ is the pinning
strength {\it relative} to the thermal fluctuation strength. The
Abrikosov factor [30] is expressed in the form
$$
\beta_A=1 + N_v^{-1} \sum_{\bf k} v_{\bf k} ( \, 1 - 2\, x \, V_{\bf
k} \, ), \eqno(2.6)$$
where $V_{\bf k}$ is the fully-renormalized four-point vertex
corresponding to the bare one $v_{\bf k}$ and also depends on
$\Delta_{\rm eff}$. If the freezing to the solid is of first order, a
precursor of the positional ordering of vortices (zero points of $\psi$)
will appear only in the vicinity of the transiton. Then, sufficiently
above the transition, $V_{\bf k}$ will take a form of RPA type such as
$$
V^{\rm liq}_{\bf k} \simeq {{v_{\bf k}} \over {1 + 2 \, x \, v_{\bf
k}}}. \eqno(2.7)$$
On the other hand, one will notice by comparing eq.(2.6) with its mean
field
expression [30] that $V_{\bf k}$ in the limit of a perfect solid takes
the form $$
V^{\rm sol}_{\bf k}=V^{\rm liq}_{\bf k}(x \gg 1) + \delta V_{\bf k}
=
{1 \over {2 x}} ( \, 1 - N_v \sum_{{\bf G} \neq 0} \delta_{{\bf k}, {\bf
G}} \, ), \eqno(2.8)$$
where ${\bf G}$'s are the reciprocal lattice vectors of the vortex
solid. Note that the first term corresponds to the $x \to \infty$ (i.e.,
low $T$) limit of $V^{\rm liq}_{\bf k}$, while the second term $\delta
V_k$
corresponds to the structure factor of a vortex state. In a solid-like
vortex
state with a long but finite positional correlation length, $\delta
V_{\bf k}$
near ${\bf G}$ will take, for instance, a Gaussian form like
$$
\delta V_{\bf k} \sim - {1 \over {2 \, x \, \epsilon}} \exp{\biggl(-
{{({\bf k} - {\bf G})^2}
\over {2 \epsilon h}} \biggr)}, \eqno(2.9)$$
where $\epsilon^{-1}$ ($\gg 1$) corresponds to the positional
correlation
area $N_{\rm cor}$. One can verify by substituting eq.(2.9) into
eq.(2.6)
that the $\beta_A$-value
does not depend remarkably on $N_{\rm cor}$ at least at low enough $T$.
In addition, we note that, at low $T$, eq.(2.5) reduces to the mean
field
result
$$
{\overline {< < |\psi|^2 > >_{\rm sp}}}=- (b \, \beta_A)^{-1} \,
\mu_0,
\eqno(2.10)$$
where $< \,\,\,\,\,>_{\rm sp}$ denotes space average, and
the small O($\Delta_{\rm eff}$) correction to $\beta_A$ was neglected.

Now, let us turn to the glass susceptibility in 2D, which is expressed
within
LLL in the form
$$
\chi_{\rm G}=N_v^{-1} ({\overline {< |\phi_p|^2 >}})^{-2} \sum_{p,
p'} {\overline {|< \phi_p \, \phi^*_{p'} >|^2}}. \eqno(2.11)$$
In clean limit, $\chi_G$ is given as a ladder-series of the irreducible
vertex represented in Fig.1, i.e.,
$$
\chi_G=1 + I_{\rm irr} + I_{\rm irr}^2 + I_{\rm irr}^3 + \cdot \cdot
\cdot \cdot \cdot, \eqno(2.12)$$
where
$$
I_{\rm irr}={{\Delta \, h} \over {2 \pi N_v \mu^2}} \sum_{\bf k} \,
v_{\bf k} ( 1 - 2 \, x \, V_{\bf k} )^2. \eqno(2.13)$$
Far above the freezing transition where $V_{\bf k}$ is dominated by
the RPA term (2.7), we obtain $I_{\rm irr}
\simeq \Delta_{\rm eff}/2(1 - O(x^{-1}))$.
Since this expression is insensitive to $B$,
one may conclude the absence [31] of 2D glass
transition even at the mean field level. However, once the {\it quantum}

superconducting fluctuation is taken into account, a glass transition
line at the mean field level can exist above a (if any) first
order transition line [32].

By contrast, when taking account of $\delta V_{\bf k}$ illustrated by
eq.(2.9) which becomes rather remarkable in the vicinity of the (if any)

first order line, one obtains a result suggestive of a glass
transition {\it induced} [17,22] by the vortex solidification (i.e., by
the first order transition). For instance, if substituting the
expression of the perfect solid eq.(2.8) into eq.(2.13), one finds
$$
I_{\rm irr}= {{\Delta \, h} \over {2 \pi \mu^2}} \, (\beta_A - 1) \,
N_v \eqno(2.14)$$
proportional to the total number $N_v$ of vortices. The factor $\beta_A
- 1$ ($> 0$) implies that a spatial variation of $|\psi|$ due to the
vortices is crucial in obtaining a glass ordering at or below the (if
any) first order line. The factor $N_v$ is a consequence of the
assumption of a perfect solid in the case
with a small but finite $\Delta$ and must be replaced by the
positional correlation area $N_{\rm cor}(\Delta_{\rm eff})$ in terms of,
say,
eq(2.9). Note that the origin of the large factor $N_v$ or $N_{\rm cor}$
is
the vertex correction to the impurity line (the semicircles in Fig.1).
If, as an estimation of $N_{\rm cor}$, identifying it with the
correlation area resulting from the collective pinning theory [33], we
find $N_{\rm cor} \simeq (\Delta_{\rm eff})^{-1}$. Further, since the
mean
field glass transition line $T_{\rm G}^{\rm mf}(B)$ in clean limit will
lie
just below the freezing transition, or crossover, line $T_m^{(2d)}(B)$
in 2D LLL, $\mu^2$ in eq.(2.13) may be replaced by its value at
$T_m^{(2d)}(B)$ $10^{-2} h \, b \, k_{\rm B} T (2 \pi \xi_0^2 d)^{-1}$.
Then, since the
$\beta_A$-value is insensitive to material parameters, the resulting
$I_{\rm irr}$ is almost independent of
material and physical parameters. At least, it does {\it not}
vanish in clean limit ($\Delta_{\rm eff} \to 0$), implying that the
disorder-free theory cannot be used even in clean limit below the
expected
$T_m^{(2d)}(B)$.

Next, let us extend the above analysis to 3D
 case in which the glass transition will occur more
easily. For a while, the case of {\it point} disorder will be
considered in which the random potentials satisfy
${\overline {u_j({\bf r}) u_l({\bf r}')}}=d^{-1} \Delta^{(p)}
\delta^{(2)}({\bf r} - {\bf r}') \, \delta_{j, l}$ 
and ${\overline {f_j({\bf r}) f_l({\bf r}')}}=d^{-1}
\Delta_\Phi^{(p)} \delta^{(2)}({\bf r} - {\bf r}')
\, \delta_{j, l}$. Roughly speaking, $\chi_G$ in 3D case is given by
replacing $\Delta/\mu^2$ in eq.(2.13) by $\Delta^{(p)}/\mu^{3/2}$ when,
for simplicity, neglecting a $\Delta_\Phi$ term. If, as in 2D case,
$N_{\rm cor}$ is identified with the dimensionless positional
correlation area perpendicular to ${\bf B}$ found in the collective
pinning theory (in type II limit) [33], we obtain
$$
I_{\rm irr} \simeq \Delta_{\rm eff} (\beta_A - 1) \exp(c_1 \,
\Delta_{\rm eff}^{-1}) \eqno(2.15)$$
along the first order line expected in LLL (see eq.(2.17) below), where
$c_1$ is a positive constant. Since eq.(2.15) is divergent in
$\Delta_{\rm eff} \to 0$
limit, it implies that the glass transition due to point
disorder, the so-called vortex-glass (VG) transition, will occur more
suddenly in cleaner systems with smaller $\Delta_{\rm eff}$ and that the
first order transition in clean limit should be a glass transition
simultaneously. Further, the exponential $\Delta_{\rm eff}$-dependence
of $N_{\rm cor}$ suggests that this pinning dependence is stronger than
the prefactor $\Delta_{\rm eff}$ in eq.(2.15). Since $N_{\rm cor}$ will
decrease down to a constant of order unity with increasing $\Delta_{\rm
eff}$, and $\beta_A$ depends only weakly on $\Delta_{\rm eff}$, it is
expected that $I_{\rm irr}$ monotonically {\it decreases} with
increasing $\Delta_{\rm eff}$ down to a value $\simeq \Delta^{(p)} h
(\beta_A-1)/(2 \pi \mu^{3/2})$. Namely, the resulting VG transition line
is expected to deviate from the first order line to {\it lower}
temperature with increasing
$\Delta_{\rm eff}(T)$ and, in high $\Delta_{\rm eff}(T)$ limit, approach
$$
B_{\rm VG}^\infty(T) \simeq H_{c2}(0) \biggl({{-\mu_0}
\over {\theta_f(T)}} \biggr)^{3/2} (\Delta^{(p)})^{1/2}. \eqno(2.16)$$
Although this line [16] satisfies the LLL scaling $B \sim (T_c(B) -
T)^{3/2}$
as well as the first order line [34] in 3D and LLL
$$
B_m(T) \simeq H_{c2}(0) {{(-\mu_0)^{3/2}} \over {\theta_f(T)}},
\eqno(2.17)$$ the dependences on the (anisotropic) 
3D fluctuation strength $\theta_f=b k_{\rm B} T
\sqrt{\Gamma}/\xi_0^3$ of $B_{\rm VG}^\infty$ and $B_m$ are different
from each other.

Through the above findings on the glass transition at or below $B_m(T)$,
a
picture on the phase diagram above a lower critical point $B_{\rm lcp}$
(defined below) of the thermal first order transition is easily obtained
[17].
First, since most of the above expressions depend not on $\Delta^{(p)}$
but on
the relative pinning strength $\Delta_{\rm eff}(T) \propto 1/T$, a
cooling
along $B_m(T)$ (i.e., an increase of $B$) will imply an effective
enhancement
of random pinning effect. Further, as mentioned below eq.(2.2),
an inclusion of nonzero $\Delta_\Phi$ additionally induces a pinning
effect
enhanced by an increase of $B$ [18,19]. For these reasons, the first
order transition along $B_m(T)$ will be weakened with
increasing $B$ and will disappear at some upper critical point $B_{\rm
ucp}$.
Then, it is clear that the vortex state just below the thermal first
order
transition and below $B_{\rm ucp}$ need not have a positional order.
Actually, as explained above, the glass transition line begins to
deviate
from $B_m(T)$ to lower temperature with increasing $\Delta_{\rm eff}(T)$
or
$B$ and approaches
$B_{\rm VG}^\infty(T)$ at high enough fields. Note that $B_{\rm
VG}^\infty(T)$
decreases with reducing the pinning strength and, as suggested below
eq.(2.17), decreases more rapidly than $B_m(T)$ with increasing
$\theta_f$. Hence,
there is a possibility [17] of a wider window of the so-called vortex
slush regime [35] in cleaner systems as far as it is not masked by the
BrG phase which may exist at lower fields (see Fig.3 below).

Of course, the vortex slush regime is a part of the vortex liquid region
and hence, of the normal metal phase because the linear resistance is
finite there. In the present theory, the in-plane resistivity
$\rho_{xx}$ ($= \rho_{yy}$) in
this regime vanishes algebraically on approacing the glass transition,
as in the strong disordered case, but with a field-dependent smaller
exponent [18].
To show this, let us briefly explain how to evaluate the conductivity
near
a VG transition. As accepted even through the studies [6] at higher
temperatures, the conductivity $\sigma_{ij}$ may be
separated into the quasiparticle part $\sigma_{n, \, ij}$ and the
superconducting fluctuation part $\sigma_{s, \, ij}$, and, deep in the
liquid
regime of clean systems, $\sigma_{s, \, ij}$ can be expressed as a sum
of
the pinning-free contribution $\sigma_{F, ij}$ and the glass fluctuation
part
$\sigma_{G, ij}$ so that $\sigma_{ij} \simeq \sigma_{n, \, ij} +
\sigma_{F, ij} + \sigma_{G, ij}$ [17].
The dynamics of $\psi$-field is incorporated according to the TDGL
equatioin,
or equivalently the quantum TDGL action [17,36]
$$
{{S_{\rm QLD}} \over \hbar}=d \sum_j \int d^2r \beta \sum_\omega
(\gamma_1 |\omega| + {\rm i} \gamma_2 \omega ) |\psi_j({\bf r},
\omega)|^2 +
\int_0^{\hbar \beta} {{d\tau} \over \hbar} \biggl[ {\cal H}_{\rm
LD}(\psi \to \psi(\tau))$$
$$ + {\cal H}_{rp}(\psi \to \psi(\tau)) \, \biggr], \eqno(2.18)
$$
for the LD model,
where $\beta=1/(k_{\rm B} T)$, $\gamma_1 > 0$,
and $\omega$ denotes Matsubara frequency. As first
found in Ref.37, the vortex flow conductivities
$$
\sigma_{F, xx}=R_q^{-1} {{\gamma_1 \, \phi_0} \over {b \, B}} (-
\mu_0) \eqno(2.19)$$
and $\sigma_{F, xy}=\gamma_2 \sigma_{F, xx}/\gamma_1$
are obtained as the low $T$ limit of the renormalized Aslamasov-Larkin
(AL)
fluctuation conductivities [38], where $R_q$ is the resistance quantum
$\pi \hbar/2 e^2$. According to eq.(4.1) of ref.6, the corresponding
diagonal AL conductivity parallel to ${\bf B}$ has the following low $T$
form obeying the LLL scaling
$$
\sigma_{F, zz} \simeq \sigma_{F, xx} \biggl(-{{2 \pi \xi_0^2 \, r_B \,
\mu_0}
\over {b k_{\rm B} T \Gamma}} \biggr)^2 \propto {{(T_c(B) - T)^3} \over
{(B
k_{\rm B} T \Gamma)^2}}. \eqno(2.20)$$
The same relation was derived later in ref.39 in terms of the phase-only
model. It suggests that the behavior (2.20) of $\rho_{zz}$ is also valid
in the liquid regime in lower fields.

Below, we
focus on the glass fluctuation term $\sigma_{G, xx}$. The Feynman
diagrams
expressing $\sigma_{G, xx}$ are illustrated in Fig.2. In weak enough
pinning
case, the resistive vanishing can be described just by Fig.2 (a), while
Fig.2 (b) becomes necessary in order to derive the universal VG scaling.

Actually, assuming the glass transition to be continuous, Fig.2 (a)
gives [17]
$$
\sigma_{G, xx}^{(a)} \sim (T-T_{\rm VG})^{(3-z)\nu}, \eqno(2.21)$$
while the diagrams such as Fig.2 (b) result in the scaling behavior [20]

argued by Fisher et al.
$$
\sigma_{G, xx}^{(b)} \sim (T - T_{\rm VG})^{(1-z)\nu}, \eqno(2.22)
$$
where $T_{\rm VG}(B)$ is the VG transition line (in type II limit), and
$z$ ($> 4$) and $\nu$ are, respectively, dynamical exponent and the
exponent
of correlation length $\xi_{\rm VG}(T)$ in VG critical region (in type
II
limit). $\sigma_{G, xx}$ is given by a sum of eqs.(2.21) and (2.22).
A crossover between the behaviors (2.21) and (2.22) occurs when [18]
$$
\xi_{\rm VG}(T) \sim L_{\rm cr}(B)=\sqrt{{{\phi_0}
\over {B \Delta_{\rm eff}(T)}} \biggl( 1 + {{\Delta^{(p)}} \over {4 h^2
\Delta_\Phi}} \biggr)}. \eqno(2.23)$$
Namely, if $L_{\rm cr}(B)$ is beyond the system size,
the behavior (2.21) may be seen like a true critical behavior
in type II limit in cleaner systems, and, in a clean sample, the
exponent of vanishing resistivity is $B$-dependent and will increase
from $\nu(z-3)$ to $\nu(z-1)$ with increasing $B$ in an apparently
continuous manner [18].
In fact, a $B$-dependent exponent was
observed in the vortex slush regime of a YBCO sample, and the expected
universal behavior (2.22) seems to have been found in higher fields than
$B_{\rm ucp}$ [40]. A similar behavior was observed previously in other
experiments: the resistivity exponent in a moderately disordered sample
was $B$-dependent and smaller than the expected one (2.22) [41], while a

$B$-independent scaling behavior has been observed in dirtier
samples [42]. This expectation [18]
of an algebraic and $B$-dependent scaling of
resistance in the vortex slush regime is different from the
thermally-activated vanishing in the slush regime argued by Worthington
et
al. [35].

So far, our discussion has been limited to the field range in which the
VG
transition occurs (at or) below the first order line or its
extraporation to higher fields than $B_{\rm ucp}$. Since, as mentioned
above, the pinning disorder
becomes effectively weaker in lower fields, it may be natural to expect
the
first order transiiton not to terminate at lower fields.
However, the {\it thermal} first order transition should not occur any
longer in fields where a glass transition line exists above $T_m(B)$,
and a lower critical point $B_{\rm lcp}$ of the first order line should
appear if the glass transition lies above $B_m(T)$ in lower fields [19].

Actually, this situation is realized even at weak disorder in the
present LLL approach which may be qualitatively valid far above the
$H_{c1}(T)$. To see this, let us first consider the case with only
{\it line}-like disorder parallel to ${\bf B}$ [19,43],
defined by ${\overline {u_j({\bf r}) u_l({\bf r}')}}=\Delta^{(l)}
\delta^{(2)}({\bf r}-{\bf r}')$ and a similar one for the $f$-potential.
The resulting glass transition due to such correlated defects $\parallel
{\bf B}$ is called in the literature the Bose-glass (BG) transition
[44].
Assuming the BG transition to occur above $B_m(T)$, the $\delta V_{\bf
k}$-contribution to the vertex correction to the impurity line, carrying
$\Delta^{(l)}$ in this case, in Fig.1 can be neglected, and the vertex
correction consists only of the RPA term (a 3D version of eq.(2.7)).
Then, $I_{\rm irr}$ is easily obtained in the lowest order in
$\Delta^{(l)}$, and a BG-line results in.
When $l_{ph}=\xi_0/\sqrt{\Gamma \mu}$ (the usual phase coherence
length $\parallel {\bf B}$) $\gg d$, it is expressed as
$$
B_{\rm BG}(T) \simeq H_{c2}(0) {{\Delta^{(l)}} \over {(\theta_f(T))^2}}
(-\mu_0), \eqno(2.24)$$
which is linear in $T_{c0} - T$ near $T_{c0}$, as observed in ref.15.
Since $B_m(T)$ (2.17) has a vanishing
curvature near $T_{c0}$, one finds that the resistivities vanish on
cooling {\it continuously} on $B_{\rm BG}(T)$ above $B_m(T)$ in $B <
B_{\rm lcp}^{(l)}$, where
$$
B_{\rm lcp}^{(l)} \simeq  {{(\Delta^{(l)})^3} \over {(\theta_f(T))^4}}
\, H_{c2}(0), \eqno(2.25) $$
increasing with reducing the fluctuation or enhancing the pinning.
Accompanying this BG transition, not only the diagonal conductivities
$\sigma_{xx}$ and $\sigma_{zz}$ but also the tilt modulus (the
diamagnetic susceptibility to a transverse field $\perp {\bf B}$) show a
critical divergence, implying the presence of a transverse Meissner
effect in the BG phase [44].
The details of calculations of these response
quantities near a BG transition will not be given here and can be found
in ref.19 together with the corresponding results in the Gaussian
splayed-glass
transitions [45].

Interestingly, the corresponding situation with $B_{\rm VG}(T) > B_m(T)$
also occurs in the case [24] with only point disorder. Under the same
assumption as that used above for $B_{\rm BG}(T)$, we find a $B_{\rm
VG}(T)$-line [17]
$$
B_{\rm VG}(T) \simeq \biggl({{\Delta^{(p)}} \over {\theta_f(T)}}
\biggr)^2 B_{dc}(T)=H_{c2}(0) {{\xi_0 (\Delta^{(p)})^2} \over
{\sqrt{\Gamma} (\theta_f(T))^3}} (-\mu_0) \eqno(2.26)$$
linear in $-\mu_0$, where $B_{dc}(T)$ is the so-called decoupling
crossover
line [34].
In this case, the resulting lower critical point $B_{\rm lcp}^{(p)}$
$$
B_{\rm lcp}^{(p)} \sim 10^{-2} H_{c2}(0) \biggl({{\xi_0} \over
{\sqrt{\Gamma} d}} \biggr)^3 {{(\Delta^{(p)})^6} \over
{(\theta_f(T))^7}} \eqno(2.27)$$
is usually much smaller than but has similar dependences on the pinning
and fluctuation strengths to the corresponding $B_{\rm lcp}^{(l)}$. We
note that the above expression was derived in the lowest order in
$\Delta^{(p)}$. An inclusion of the next order contribution to $I_{\rm
irr}$ tends to increase $B_{\rm lcp}^{(p)}$-value, although the
resulting dependences on the pinning and fluctuation strengths become
complicated [46]. Hence, eq.(2.27) should be seen as a lower limit of
the expected lower critical point. In any case, the resistivities in $B
< B_{\rm lcp}^{(p)}$ are expected to vanish continuously at the second
order VG transition, and the {\it thermal} first order transition should
not occur in these low fields where $B_m(T)$ lies in the glass phase. It
is important to note that, as clear from the above discussion, a lower
critical point is not due to an enhancement of pinning effect
accompanying a lowering of the field: A lower critical point was
observed in a couple of experiments [7-9] in tesla range of YBCO clean
samples where the type II limit neglecting fluctuations of flux density
is safely valid. It is theoretically difficult to expect a
pinning-enhancement accompanying a field-lowering in type II limit.
Actually, for this reason, the appearance of a lower critical point was
not predicted from treatments based on the vortex elasticity.

As a test of the present explanation [17,24] on the existence
of the vortex slush
regime and of $B_{\rm lcp}$, it is interesting to compare the above
results with the oxygen-deficiency dependence [9] of YBCO phase diagram.
As systematically examined by Nishizaki et al., {\it both} the upper and
lower critical points of first order line tend to decrease with
underdoping [9,10]. It is well known at present through the doping
dependences of penetration depth [47] and heat capacity jump [48] that
the thermal superconducting fluctuation is enhanced with underdoping. On
the other hand, effects of point disorder due to oxygen deficiency
also become more remarkable with underdoping. Namely, both
$\theta_f(T_{c0})$ and $\Delta^{(p)}$ increase with underdoping. It is
natural to interpret the doping dependence of $B_{\rm ucp}$ as being due
to a $\Delta_{\rm eff}(T)=\Delta^{(p)}/\theta_f(T)$-increase (i.e., a
relative enhancement of
pinning) with underdoping. However, additional dependences on the
anisotropy $\Gamma$ and on $\theta_f$ of $B_{\rm lcp}^{(p)}$ (2.27)
imply that $B_{\rm lcp}^{(p)}$ can decrease with increasing $\Delta_{\rm
eff}$, because $\Gamma$ remarkably increases with
underdoping [48]. An explanation of other experimental findings on the
lower critical points was given in ref.24.
Now, we are in a position of discussing possible phase diagrams under
${\bf B} \perp$ layers, which are described in Fig.3 [24]. Before
explaining Fig.3, we need to mention a bit about results of the elastic
approach. As emphasized in, for instance, ref.12, the melting line of
the BrG phase should be of first order in general. However, the
superconducting transition observed
in $B < B_{\rm lcp}$ is of second order, and it is difficult to explain
the presence of $B_{\rm lcp}$ consistently with the argument favoring
the BrG phase. As far as the lower critical point is unrelated to the
BrG melting, there is no reason why all of the thermal first order line
in $B_{\rm lcp} < B < B_{\rm ucp}$ is included in the BrG melting line.
Through some consideration including the above statements, we have
concluded that a generic 3D phase diagram in ${\bf B} \perp$ layers with
intermediate strengths of fluctuation and pinning will be of the form
Fig.3 (a), in which the BrG melting is separated from the thermal first
order transition occuring along $B_m(T)$, and the vortex slush regime
exists entirely in $B_{\rm lcp} < B <  B_{\rm ucp}$. An evidence of a
BrG melting line lying much below $B_{\rm VG}(T)$ was recently found in
data of NbSe$_2$ [49] and (K, Ba)BiO$_3$ [50]. However, in HTSC with
strong fluctuation, the thermal first order line may be pushed down to
lower temperature and, in part, merge the BrG melting line. Then, the
only possible phase diagram will be of the form Fig.3 (b), in which the
glass phase just below the first order transition in $B_{\rm lcp} < B <
B^*$ is BrG, and the vortex slush regime exists only in $B^* < B <
B_{\rm ucp}$. A recent magnetization measurement in heavily overdoped
YBCO [51] seems to have shown the presence of the BrG melting line in $B
< B_{\rm lcp}$ approaching $T_{c0}$ with decreasing $B$ just like what
we expect through Fig.3 (b). The dashed curves in both Fig.3 (a) and (b)
indicate $B^\infty_{\rm VG}(T)$. In passing, we note that the phase
diagrams given in Fig.3 are valid even for real systems including weak
line-like disorder, although in this case the glass phases have the
transverse Meissner effect, and the vortex slush regime is much narrower
[24] than in the case with no line disorder.

\vspace{8mm}
\leftline{\bf 3. PHASE DIAGRAM IN FIELDS PARALLEL TO LAYERS}
\vspace{5mm}

In this section, we briefly explain results found in LLL approach to the
phase
diagram of model (1.1) in ${\bf B} \parallel$ layers. In this field
configuration, the field strength is usually measured by the
combination [26,52]
$$p \equiv {{2 \pi d^2} \over {\phi_0}} \sqrt{\Gamma} B, \eqno(3.1)$$
as far as the relation $\xi_0 < \sqrt{\Gamma} d/\sqrt{2}$ is satisfied.
Below, we primarily focus on the strong field region satisfying
$\exp(-p) \ll 1$ in such a layered system.
In such high fields, the mean field transition line $T_c(B)$ approaches
a limiting behavior [53]
$$
T_c(B) \to T_{c0} \biggl( 1 - {{2 \xi_0^2} \over {\Gamma d^2}} \biggr)
\eqno(3.2)$$
independent of $p$, and the action (2.18) written in terms of LLL modes
takes the simple form [26,27]
$$
{{S_{\rm QLD}} \over \hbar}=\beta \sum_{\bf Q} \sum_\omega \,
\gamma_1 |\omega| |\phi_\omega({\bf Q})|^2 + \int_0^{\hbar \beta}
{{d\tau} \over \hbar} \sum_{\bf Q} \biggl[ \, \biggl( \mu_0 + \xi_0^2
\sum_{\mu=x, y} \biggl( q_\mu + {{2 \pi} \over {\phi_0}} \delta
A_\mu(\tau) \biggr)^2 \biggr)$$
$$\times |\phi({\bf Q}, \tau)|^2 + {b \over {2 d L^2}} \sum_{{\bf Q}_1,
{\bf Q}_2, {\bf Q}_3} V_0(n_1-n_3, n_2-n_3; q_{y, i})$$
$$\times \phi^*({\bf Q}_1, \tau) \phi^*({\bf Q}_2, \tau) \phi({\bf Q}_3,
\tau) \phi({\bf Q}_1+{\bf Q}_2-{\bf Q}_3, \tau) \, \biggr], \eqno(3.3)
$$
where $\phi_\omega$ is the Fourier transform of the LLL fluctuation
field
$\phi(\tau)$, ${\bf Q}_j=q_x {\hat x} + Q_j {\hat y}$, $Q_j==
q_{y, j} +
r_B^{-2} d \, n_j$ with integer $n_j$,
$\mu_0=(T - T_c(B))/T_{c0}$ with eq.(3.2),
and the disorder energy term ${\cal H}_{rp}$ was dropped
for convenience of our presentation. The bare vertex $V_0$ is an even
function
[26] of $n_1-n_3$ and $n_2-n_3$, implying that the partial LLL
degeneracy of degree $N_d=L/d$ is measured by $n_j$.
Further, the gauge disturbance $\delta {\bf A}$ necessary in deriving a
conductivity at a uniform current was assumed to be
spatially uniform. The corresponding action useful in deriving tilt
modulus
can be seen in ref.27.

In the {\it disorder-free} case, the only true transition is argued
again to be a first order transition at $T_m(B)$
between a vortex solid and a (narrow) vortex liquid regime below
$T_c(B)$ [26]. In low fields, $T_m(B)$ is close to the corresponding one

of the anisotropic 3D
 GL model, which is given by eq.(2.17) with $\Gamma$ replaced by $1$,
while it approaches, as well as $T_c(B)$, a $p$-independent value in
large
$p$ limit [26,27]:
$$
T_m(p \gg 1) \simeq {T_c(B)} \biggl( 1 + c_m \, \theta_f^{(2d)}(T_{c0})
\biggr)^{-1} \eqno(3.4)$$
with eq.(3.2), where $\theta_f^{(2d)}(T)=b \, k_{\rm B} T/(2 \pi
\xi_0^2 d)$
is the fluctuation strength in 2D, and a constant of order unity $c_m$
($> 0$)
has not been
determined analytically. The fact that $T_m(B)$ becomes independent of
$p$ for
high $p$ values is a reflection of confinement of vortices between all
interlayer spacings. Because an increase of $p$ in $p > 1$ does not
delocalize the vortices out of interlayer spacing any longer but just
compresses each vortex row along the
layers, the spatial variation of $|\psi|$ on the superconducting layers
diminishes with increasing $p$. Namely, since the Abrikosov factor
$\beta_A$, eq.(2.4), approaches $1.0$ with increasing $p$ ($> 1$)
irrespective of the vortex lattice structure at lower temperatures, the
first order transition becomes significantly weaker with increasing $p$.
In the simulation of 3D XY
model [54] where $T_c(B)$ is always identical with $T_{c0}$, a similar
melting
line insensitive to $p$ in $p > 1$ was detected from heat capacity data.

However, the high $p$ ($> 1$) portion of the melting transition was
argued
there not to be weakened first order mentioned above but to be
continuous.
This controversy may not be resolved by real experiments because, as
discussed below, there is a reason why the disorder-free melting
transition
in higher $p$ should be easily destroyed by disorder existing in real
systems
[27]. Here we merely note that $T_m$ of eq.(3.4) decreases with
increasing the fluctuation strength $\theta_f^{(2d)}(T_{c0})$ or with
{\it decreasing} the anisotropy $\Gamma$. The latter dependence seems to
become dominant in the doping dependence in YBCO according to
resistivity data in ref.55.

Below $T_m(B)$ of the {\it disorder-free} system, a
Josephson-vortex-solid, i.e., a solid phase pinned by the layer
structure, is created. This low $T$ phase has finite helicity
moduli and thus, zero resistance for any direction on the layer, while
it
has a nonzero vortex flow resistance guaranteed by zero helicity modulus
in
the perpendicular direction to the layers.
Consequently, there is a transverse Meissner effect for a tilt
perpendicular to the layers but not for any tilt parallel to the layers.
With a slight change of the $p$-value, a structural transition possibly
mediated by a
{\it unpinned} solid occurs between different pinned solids and is
reflected
on the $T_m(B)$ line in intermediated fields as its oscillating
$B$-dependence [26]. This oscillating behavior has been suggested in
YBCO data [55,56].
However, a description of $p$-dependences of such consecutive
structure transitions is highly complicated and will not be given here.

The isotropic form of the gradient terms in eq.(3.3) within the layers
(in $x$-$y$ plane) leads to a key insight on the physical picture in the
liquid regime. It implies that, for high enough $p$ values, the linear
responses, such as the resistance, measured along the layers are
independent of the relative direction between $\delta {\bf A}$ (i.e.,
the current)
and ${\bf B}$. This is the
essence of the so-called in-plane Lorentz force-free behavior observed
[57,58]
in tesla range of BSCCO where $p > 1$ is safely valid. Simultaneously,
the in-plane isotropic form of gradients implies that, on cooling, the
phase coherence
lengths grow isotropically on the layers, while, as in the case ${\bf B}
\perp$ layers [34], the phase coherence perpendicular to the layers
above $T_m$
is not sensitive to cooling and remains microscopic as a result of the
(partial) LLL degeneracy. Since the phase correlation must be compatible
with the positional correlation of vortices [25,26], the above-mentioned
anisotropy appearing in the phase correlation must be also satisfied by
the positional correlation. Hence, the observed in-plane Lorentz-force
free behavior proves that, above $T_m$, the positional correlation first
grows along the layers rather than across the layers, which is an
opposite trend to an argument favoring a vortex smectic liquid [59].
Namely, the observation [57,58] is incompatible
with assuming the intermediate phase [59].

It is not easy to describe in details the glass transition due to point
disorder in this case ${\bf B} \parallel$ layers. In ref.27, a high
field behavior of
glass transition line, corresponding to eq.(2.16) in ${\bf B} \perp$
layers,
was found and its interpolated behavior to lower fields was conjectured.

Applying a similar analysis to that leading to eq.(2.14) to the present
case,
the irreducible vertex $I_{\rm irr}$ of $\chi_G=(1 - I_{\rm
irr})^{-1}$, 
when $N_{\rm cor} \sim$ O(1) and $e^{-p} \ll 1$, is found to have the
form
$$
I_{\rm irr} \simeq {\Delta \over {2 \pi \mu}} \exp(-2 p), \eqno(3.5)$$
where $\mu$ in the present case satisfies
$$
\mu \simeq \exp( \, 2 \mu_0/\theta_f^{(2d)}(T) \,) \eqno(3.6)$$
which is identical with eq.(2.10), and the random flux $\Delta_\Phi$ was
again neglected. The $e^{-p}$-dependence of eq.(3.5) is a reflection of
$\beta_A -1 \sim e^{-p}$, i.e., of a weak spatial variation
of $|\psi|$ and is also related to an exponentially small shear modulus
of the pinning-free solid [60,26], while the exponential $T$-dependence
in eq.(3.6) is a reflection of a 2D-like superconducting fluctuation
weakened by a partial breaking, due to the layering, of the LLL
degeneracy. Further, the $p$-insensitive
melting line, eq.(3.4), is a consequence of eq.(3.6) which is also
independent
of $p$.

Here we will assume the first order transition line $T_m(B)$ to
terminate at some $p$ ($> 1$) and hence to have a upper critical point
indicated as $p_c$ in Fig.4.
This is reasonable, because the above-mentioned exponentially small
shear modulus in
$p > 1$ likely results in a stronger random-pinning effect with
increasing $p$. Then, even the first order transition in $p < p_c$ may
not be accompanied by
an ordinary superconducting
ordering signaled by $\mu \to 0$, and eq.(3.6) becomes valid even below
$T_m$. Consequently, using the above expressions, one obtains a
Josephson-vortex
glass (JG) transition line
$$
B_{\rm JG}(T) \simeq {{\phi_0} \over {2 \pi d \xi_0 \theta_f(T)}}
\biggl[ 1 - 2 {{\xi_0^2} \over {\Gamma d^2}} - {T \over {T_{c0}}}
\biggl(1 + {{\theta_f^{(2d)}(T_{c0})} \over 2} {\rm ln}{{2 \pi} \over
\Delta} \biggr) \biggr],
\eqno(3.7)$$
which has a similar $T$-dependence to eq.(2.26). Note that the
prefactor decreases with increasing 3D fluctuation strength $\theta_f$
($\propto \sqrt{\Gamma}$) and that, due to the assumption $N_{\rm cor}
\sim 1$,
the $\Delta \to 0$ limit cannot be taken in eq.(3.7).
In the sketched phase diagram Fig.4, eq.(3.7) is useful just
in $p > p_{c1}$. In $p_c < p < p_{c1}$, an exponential decay of $N_{\rm
cor}$
resulting from a similar $p$-dependence of shear modulus will not be
negligible with increasing $p$. Then, the prefactor of eq.(3.7)
effectively diminishes in this field range, and the resulting $B_{\rm
JG}(T)$, as described in Fig.4, becomes more flat. This phase diagram
Fig.4 will be compared with existing data in HTSC. An a.c.
susceptibility measurement for examining the onset of lock-in phenomena
(i.e., of the transverse Meissner effect) in BSCCO has been performed
[28], and the onset line of lock-in behavior found there has shown a
relation similar to eq.(3.7)
appearing in $p > p_{c1}$ in Fig.4. A detailed resistivity measurement
has been performed in an optimally-doped BSCCO above 60 (K) [29]
and has shown, at lower
temperatures, a {\it continuous} vanishing of resistance along a flat
curve
consistent with $B_{\rm JG}(T)$ of Fig.4 but, near $T_{c0}$, a
discontinuous resistivity vanishing at $B_{\rm JG}(T)$ just on or below
the disorder-free melting line $T_m(B)$. Further, recent
resistivity data in 60K YBCO have suggested the presence of a remarkable
slush
regime and of a $B_{\rm lcp} \simeq $ 7 (T) above which $T_m$ is roughly
independent of $p$ [56]. The occurrence of a $B_{\rm lcp}$ in $p \sim 1$
of the case
${\bf B} \parallel$ layers is not surprising if the system is moderately
dirty so that the VG transition line in $p < 1$ lies above the
$T_m(B)$-line. In fact, the ''vertical'' $T_m(B)$ in $p > 1$ suggests
that a situation with $B_{\rm lcp}$ and with the first order transition
at higher $p$ occurs more easily than in the case ${\bf B} \perp$
layers. The continuous
resistivity vanishing in 90K YBCO reported previously [61] might be a
phenomenon below a $B_{\rm lcp}$.  

In the case with disorder, the resistances for a current in {\it all}
directions vanish simultaneously at $T_{\rm JG}(B)$. Further, the
transverse Meissner effect signalled by a critical divergence of tilt
modulus for a
tilt {\it across}
the layers is expected to occur for any $p$-value, i.e., even if the
vortex
lattice in the ${\it pinning-free}$ case is a unpinned solid. A detailed
analysis leading to these conclusions on response properties is seen in
ref.27.

\vspace{8mm}
\leftline{\bf 4. SUMMARY AND DISCUSSION}
\vspace{5mm}

We have briefly explained the existing theory of phase diagram and
physical properties deep in the vortex liquid regime based on the LLL
approximation of
the GL model. This theory can describe the vanishing behaviors of
resistance
in various situations in a way consistent with the phase diagram, while
it is unclear to what extent the glass phases and possible transitions
{\it between}
them are described starting from the original GL model.

In $\S 2$, a comparison of the theory with experimental data was done
for
YBCO. In BSCCO with stronger fluctuation and much larger anisotropy, it
is known through the angular dependence [62] that the internal gauge
fluctuation (leading to a magnetic screening changing features of
interaction between the vortices) is no longer negligible in the low
fields where the first order transition is
realized, and hence, the present theory will not be directly applicable.

Nevertheless, we note that, due to the strong dependence on the
anisotropy of $B_{\rm lcp}^{(p)}$ in eq.(2.27), the absence of a lower
critical point in BSCCO
is not surprising. The position of the upper critical point and the
presence of slush regime in BSCCO are not clear yet to us. We simply
guess here that, like in Fig.4, the VG transition curve and its high
field limit (corresponding to eq.(2.16)) in BSCCO should lie extremely
below an extrapolated curve of $T_m(B)$ to higher fields.

Recently, the presence of pinning-induced growth of $|\sigma_{xy}|$ near
a glass transition was shown theoretically [63] and experimentally
[64,65]
and cannot be explained correctly based on the mean field vortex
dynamics neglecting thermal fluctuation [66]. This should be the case,
because the present extension of the
fluctuation theory [6] has explained the details of phase transition
lines at which the linear dissipation vanishes. The glass fluctuation
contribution $\sigma_{G, xy}$ to the Hall conductivity is expressed, as
well as $\sigma_{G, xx}$, by Fig.2, and it is expected that the
magnitude of $\sigma_{G, xy}$ grows on approaching a glass transition
with its sign, relative to that of $\sigma_{F, xy}$, dependent on the
dimensionality of a dominant pinning disorder [63]. The phenomena in the
case with only line disorder are the best understood ones, and we find
for this case that $\sigma_{G, xy} \cdot \sigma_{F, xy} < 0$ [63,65] and
that the ratio $\sigma_{G,xy}/\sigma_{G,xx}$ (i.e., the Hall angle near
the transition) does not vanish but
seems to approach a constant at the transition [65,66].
Its details will be reported elsewhere.

\vspace{16mm}
{\leftline {\bf References}}
\vspace{5mm}
\parindent=0mm

[1] H. Safar et al., Phys. Rev. Lett. {\bf 69} (1992) 824.

[2] W. K. Kwok et al., Phys. Rev. Lett. {\bf 69} (1992) 3370.

[3] U. Welp et al., Phys. Rev. Lett. {\bf 67} (1991) 3180.

[4] E. Zeldov et al., Nature (London) {\bf 375} (1995) 373.

[5] See, for instance, S. Sarti et al., Phys. Rev. {\bf B 56} (1997)
2356.

[6] R. Ikeda, T. Ohmi and T. Tsuneto, J. Phys. Soc. Jpn. {\bf 60} (1991)
1051.

[7] W. K. Kwok et al., Phys. Rev. Lett. {\bf 84} (2000) 3706.

[8] L.M. Paulius et al., Phys. Rev. {\bf B 61} (2000) R11910.

[9] T. Nishizaki et al., Physica (Amsterdam) {\bf C 341-348} (2000) 957.

[10] T. Nishizaki et al., J. Low Temp. Phys. {\bf 117} (1999) 1375; K.
Shibata et al., unpublished. 

[11] T. Giamarchi and P.le Doussal, Phys. Rev. {\bf B 55} (1997) 6577.

[12] T. Nattermann and S. Scheidl, Adv. Phys. {\bf 49} (2000) 607.

[13] D. Ertas and D.R. Nelson, Physica (Amsterdam) {\bf C 272} (1996)
79.

[14] E. Brezin, D.R. Nelson, and A. Thiville, Phys. Rev. {\bf B 31}
(1984)
7124.

[15] A.W. Smith et al., Phys. Rev. {\bf B 59} (1999) R11665.

[16] A.T. Dorsey, M. Huang, and M.P.A. Fisher, Phys. Rev. {\bf B 45}
(1992) 523.

[17] R. Ikeda, J. Phys. Soc. Jpn. {\bf 65} (1996) 3998.

[18] R. Ikeda, J. Phys. Soc. Jpn. {\bf 66} (1997) 1603.

[19] R. Ikeda, J. Phys. Soc. Jpn. {\bf 69} (2000) 559.

[20] D.S. Fisher, M.P.A. Fisher, and D.A. Huse, Phys. Rev. {\bf B 43}
(1991) 130.

[21] G. Blatter et al., Rev. Mod. Phys. {\bf 66} (1995) 1125.

[22] R. Ikeda, J. Phys. Soc. Jpn. {\bf 65} (1996) 1170.

[23] W.E. Lawrence and S. Doniach, {\it Proc. 12th Int. Conf. on Low
Temp.
Phys.}, Kyoto, 1970, ed. by E. Kanda (Keigaku, Tokyo, 1971) page 361.

[24] R. Ikeda, J. Phys. Soc. Jpn. {\bf 70} (2001) 219.

[25] M.A. Moore, Phys. Rev. {\bf B 45} (1992) 7336; R. Ikeda, T. Ohmi,
and T. Tsuneto, J. Phys. Soc. Jpn. {\bf 61} (1992) 254.

[26] R. Ikeda and K. Isotani, J. Phys. Soc. Jpn. {\bf 68} (1999) 599.

[27] R. Ikeda and H. Adachi, J. Phys. Soc. Jpn. {\bf 69} (2000) 2993.

[28] S. Nakaharai et al., Phys. Rev. {\bf B 61} (2000) 3270.

[29] J. Mirkovic, S.E. Savel'ev, E. Sugahara, and K. Kadowaki, preprint.
Quite
recently, this result in fields parallel to the layers was drastically
changed  by lowering the applied current [a talk in 56th Annual Meeting,
Japan Physica
Society (March, 2001)]. In a low enough current, the resistance has
vanished
{\it discontinuously} along a line insensitive to $B$ in a most field
range
examined there. This is consistent with the proposed phase diagram
[26,27]
in clean enough case but contradicts that argued [54] based on the XY
model.

[30] D. Saint-James, E.J. Thomas, and G. Sarma, {\it Type II
Superconductivity} (Pergamon press, Oxford, 1969).

[31] M. V. Feigel'man, V.B. Geshkenbein, and A.I. Larkin, Physica
(Amsterdam) {\bf C 167} (1990) 177. 

[32] H. Ishida and R. Ikeda, unpublished.

[33] A.I. Larkin and Yu. Ovchinnikov, {\it Nonequilibrium
Superconductivity}, ed. by D.N. Langenberg and A.I. Larkin
(North-Holland, Amsterdam, 1986)
$\S$ 5.3.

[34] R. Ikeda, J. Phys. Soc. Jpn. {\bf 64} (1995) 1683.

[35] T.K. Worthington et al., Phys. Rev. {\bf B 46} (1992) 11854.

[36] R. Ikeda, Int. J. Mod. Phys. {\bf B 10} (1996) 601.

[37] R. Ikeda, T. Ohmi, and T. Tsuneto, J. Phys. Soc. Jpn. {\bf 59}
(1990) 1397.

[38] L.G. Aslamasov and A.I. Larkin, Phys. Lett. {\bf A 26} (1969) 238.

[39] A.E. Koshelev, Phys. Rev. Lett. {\bf 76} (1996) 1340.

[40] T. Nishizaki et al., Phys. Rev. {\bf B 61} (2000) 3649.

[41] J.A. Fendrich et al., Phys. Rev. Lett. {\bf 74} (1995) 1210.

[42] A.M. Petrean et al., Phys. Rev. Lett. {\bf 84} (2000) 5852.

[43] R. Ikeda, J. Phys. Soc. Jpn. {\bf 68} (1999) 728.

[44] D.R. Nelson and V.M. Vinokur, Phys. Rev. {\bf B 48} (1993) 13060.

[45] T. Hwa, P. le Doussal, D.R. Nelson, and V.M. Vinokur, Phys. Rev.
Lett.
{\bf 71} (1993) 3545.

[46] R. Ikeda, unpublished.

[47] Y.J. Uemura et al., Phys. Rev. Lett. {\bf 66} (1991) 2665.

[48] J.W. Loram et al., J. Superconductivity {\bf 7} (1994) 243.

[49] Y. Paltiel et al., cond-mat/0008092.

[50] T. Klein et al., J. Low Temp. Phys. {\bf 117} (1999) 1353.

[51] T. Nishizaki, K. Shibata, and N. Kobayashi, unpublished.

[52] S.E. Korshunov and A.I. Larkin, Phys. Rev. {\bf B 46} (1992) 6395.

[53] R.A. Klemm, A. Luther, and M.R. Beasley, Phys. Rev. {\bf B 12}
(1975)
877.

[54] X. Hu and M. Tachiki, cond-mat/0003068.

[55] S.N. Gordeev et al., Phys. Rev. Lett. {\bf 85} (2000) 4594.

[56] T. Naito et al., unpublished.

[57] Y. Iye, T. Tamegai, and S. Nakamura, Physica (Amsterdam) {\bf C
174} (1991) 227.

[58] K. Kadowaki, Physica (Amsterdam) {\bf C 185-189} (1991) 1811.

[59] L. Balents and D.R. Nelson, Phys. Rev. {\bf B 52} (1995) 12951.

[60] B.I. Ivlev, N.B, Kopnin, and V.L. Pokrovsky, J. Low Temp. Phys.
{\bf 80} (1990) 187.

[61] W.K. Kwok et al., Phys. Rev. Lett. {\bf 72} (1994) 1088.

[62] S. Ooi et al., Phys. Rev. Lett. {\bf 82} (1999) 4308.

[63] R. Ikeda, Phys. Rev. Lett. {\bf 82} (1999) 3378; Physica
(Amsterdam)
{\bf C 316} (1999) 189.

[64] G. D'Anna et al., Phys. Rev. Lett. {\bf 82} (1999) 3379.

[65] W. Goeb et al., Phys. Rev. {\bf B 62} (2000) 9780; W.N. Kang et
al., Phys. Rev. {\bf B 61} (2000) 722.

[66] R. Ikeda, in preparation.

\vspace{30mm}

\begin{figure}[h]
\begin{center}
 \includegraphics{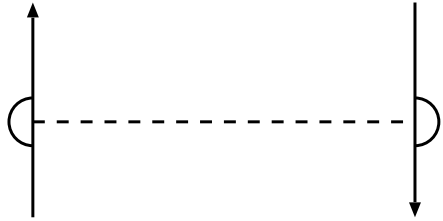}
\end{center}
 \caption{Diagram representing $I_{\rm irr}$. The straight
line denotes the LLL propagator, the dashed line is the ''impurity''
line carrying $\Delta$, and the semicircles imply vertex correction.}
\end{figure}

\begin{figure}[h]
\begin{center}
\scalebox{0.6}[0.6]{ \includegraphics{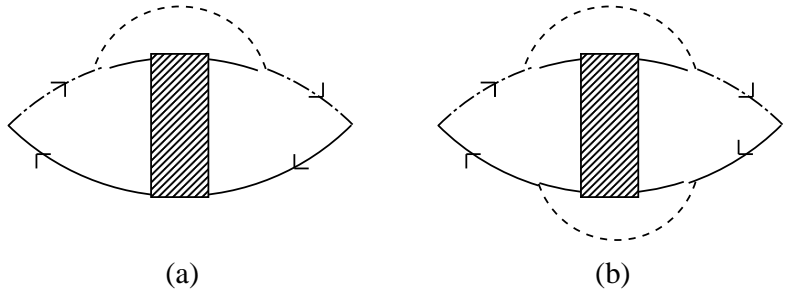}}
\end{center}
 \caption{Two typical Feynman diagrams contributing to
$\sigma_{G,xx}$, in which the chain line denotes the next lowest Landau
mode of $\psi$, and the hatched rectangle implies the correlation
function (1.1).}
\end{figure}

\begin{figure}
\begin{center}
 \includegraphics{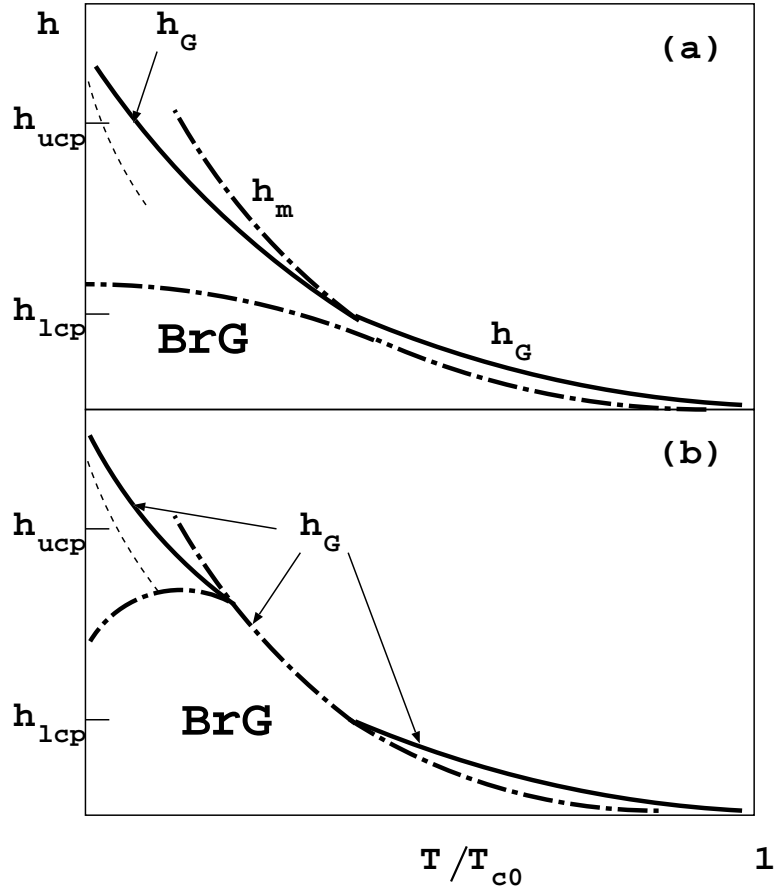}
\end{center}
 \caption{ Schematic phase diagrams conjectured for ${\bf B}
\perp $ layers. The specific case (b) follows from the generic one (a)
as a consequence of strong fluctuation. The solid curve denotes the
second order glass transition, while the chain curves include both the
thermal first order line and the BrG melting line.}
\end{figure}

\begin{figure}
\begin{center}
\rotatebox{270}{ \includegraphics{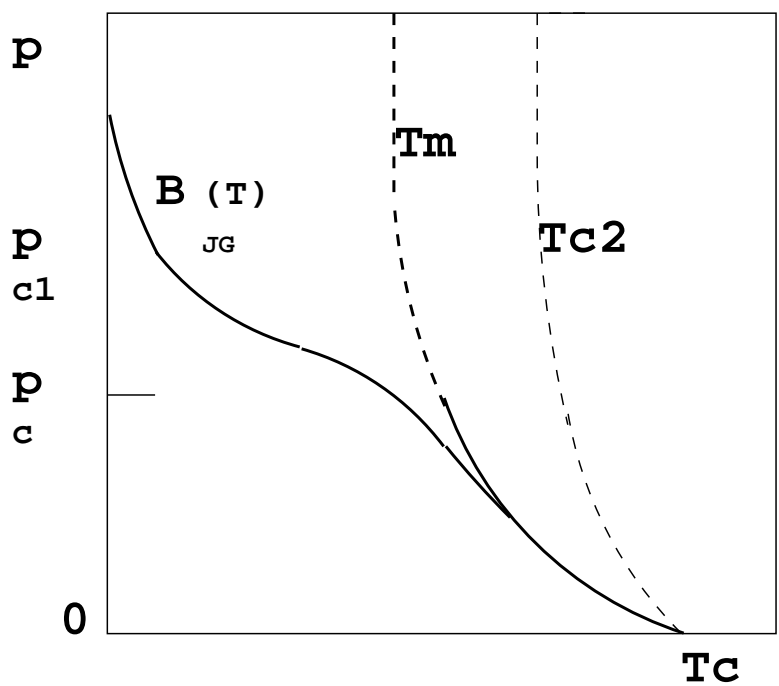}}
\end{center}
 \caption{ Schematic phase diagram for the case ${\bf B}
\parallel$ layers. $T_m(B)$ and $T_c(B)$ denote, respectively, the
melting curve in the pinning-free case and the $H_{c2}(T)$ crossover
line. The $B_{\rm JG}(T)$ curve and the first order transition line are
expressed by, respectively, the solid and chain curves. Any reflection
of possible structural transitions below $T_m(B)$ is not described here.}
\end{figure}

\end{document}